\documentstyle[11pt]{article}
\input{epsfig.sty}

\newcommand{\be}{\begin{equation}}
\newcommand{\ee}{\end{equation}}
\newcommand{\bc}{\begin{center}}
\newcommand{\ec}{\end{center}}
\newcommand{\ba}{\begin{array}}
\newcommand{\ea}{\end{array}}
\newcommand{\vx}{{\bf x}}

\newcommand{\vy}{{\bf y}}

\newcommand{\bx}{{\bf x}}

\newcommand{\by}{{\bf y}}

\newcommand{\cH}{{\cal H}}

\newcommand{\lap}{\lambda \phi^4}

\newcommand{\bphi}{{\bar{\phi}}}

\begin{document}

\begin{titlepage}

\title{
 {\bf
Free parameters in 
quantum field theories: 
an analysis of the variational approach
}
           } 
\author{  {\bf F\'abio L. Braghin }\thanks{e-mail:
braghin@if.usp.br }   \\
{\normalsize 
Instituto de F\'\i sica da Universidade de S\~ao Paulo } \\
{\normalsize C.P. 66.318,  C.E.P. 05315-970, S\~ao Paulo,    Brasil }
}

\maketitle
\begin{abstract}
The usual renormalization procedure 
for the variational 
approximation with a trial  Gaussian ansatz
for the $\lap$ model in 3+1 dimensions is re-analysed as a departing
framework for the investigation of the  parameters of the model. 
The so-called asymmetric phase of the model (where
$<vac|\phi |vac> \neq 0$) is considered
for the search of privileged values 
of  these parameters (mass and coupling constant)
and  possible conditions they may be expected to satisfy.
This also may yield a suitable approach for the investigation of the
 reliability and stability of the approximation.
The extremization of the renormalized energy density
with relation to the renormalized  mass,  coupling and 
$\bphi$ is done.
The minimizations of the renormalized energy with relation to 
the mass and $\bphi$ provide different
expressions from the ones obtained by the usual variational
procedure for the regularized theory.
Sort of ``energy scale'' invariances in  expressions for the renormalized 
mass and coupling constant  are found.
A different view on the restoration of symmetry issue  is discussed.
The transcendental character of the GAP equation
 may be reduced or even 
eliminated  by placing some variables in the complex plane.
\end{abstract}

\vskip 0.4cm



\vskip 0.3cm


\end{titlepage}

\newpage
\setcounter{equation}{0}

\section{ Introduction}

The most developped approach to solve interacting 
field (and many body) theories
is perturbation theory which only works well for very small coupling 
constants as it occurs in Quantum Electrodynamics. 
Also in this approach there is a 
systematic and direct way of 
dealing with ultraviolet (UV)
divergences, i.e.,  one knows precisely how to 
renormalize the parameters and to make the theory finite 
\cite{ITZ-ZUB,RAMOND,STEFANOVICH}. 
There are many motivations for the development of
non perturbative methods in Quantum Field and Many Body 
Theories such as for 
the description of strong interacting systems (with or without 
spontaneous symmetry breaking(s) (SSB)),
bound states and phase transitions.
One  method which has been quite extensively investigated 
is the variational
approximation which, with the use of Gaussian wave functional,
has been showed to be useful in a wide
variety of situations. 
It corresponds to  a summation of  ``cactus'' type loop diagrams 
 \cite{BAGAN,STEVENSON,BARMOSHE,KMV,DMITRASINOVIC}.
 It is equivalent
to the Hartree Bogoliubov approach \cite{THESE} and
also to 
the leading order large N approximation
\cite{COOPERetal,ZINNMOSHE}.
In this approach the ground state of the system 
is determined by equations for the variational parameters,
which are choosen to be a mass and the classical expected 
field characteristic from  a SSB state ($<\varphi> \equiv \bar{\varphi}$, 
which will be referred to as condensate 
\cite{THESE}).
These equations are derived by 
the minimization of the (regularized) averaged energy density
with respect to variational parameters of the trial wavefunctional. 
The subtraction of the (equations of the)
 theory with $<\varphi> =0$ from the ones of the 
theory in which $<\varphi> \neq 0$ provides a general and
consistent elimination of the ultraviolet divergences.
Limitations pointed out and discussed in 
\cite{FEYNMAN} have been 
rediscussed, leading to 
extensions and higher order calculations for static and 
time dependent formulations
\cite{2gaussianas,VASILEV-DAWSON,POSTGAU,KIM-YOU,HADRON,OPTetc,CEATEDESCO,COOPERetal,t-dep,t-dep2,t-dep3,LBetal,BDVetal,BOYANOVSKY,t-dep4,FLB2001PRD,ZINNMOSHE}. 
For the sake of conciseness
they will not be discussed here although the 
main ideas of the present work apply to complementary
approaches.

It is
usually highly desirable to predict the values of the
free parameters of a physical theory, 
such as masses and couplings, from the theory itself
before comparisons to experimental observations.
For this there may be possible to predict values, 
either exact ones or "privileged" range. 
These "privileged" values may also be associated to the
validity of the approximation method or even about 
applicability of the model.
The main aim of the present work is to suggest and investigate
one reasoning
according to  which  values (or range of values) 
for these parameters could be found.
Eventually this may constraint them in a more specific theory.
Basic ideas are to search for renormalized 
couplings and masses which extremize (minimize/maximize)
 the renormalized energy density.
This procedure can be considered as complementary to
the renormalization group method \cite{ZINNJUSTIN}.
Another procedure will be to consider parameters in the complex plane
to introduce auxiliar (imaginary) variables which are to be eliminated.

The $\lambda \phi^4$ model has been extensively studied for  different
reasons among which
to shed light on non perturbative effects in quantum field and many
body theories (QFT, QMBT). 
It corresponds to one of the simplest self interacting
model whose structure is expected to be (partially) present 
in several more elaborated theories and it presents interesting
features 
\cite{HUMAPO,BRANCHINA,KMV,ZINNJUSTIN,TRACEANOMALY,CONSOLI,ASYMPFREE,FROLICH}. 
It has also been considered for the study of cosmological models
\cite{COSMO} and of the Higgs particle in the standard model, 
for example in \cite{CONSOLI}.
It also shares several properties with the linear 
sigma model (LSM) which is an effective model for low energy QCD. 
Although it strongly seems to possess
asymptotic freedom in the asymmetric phase 
\cite{BRANCHINA,KMV,ASYMPFREE}, 
the model is 
``trivial'' in the symmetric phase \cite{STEVENSON,FROLICH,ZINNJUSTIN}.

In the present work the usual renormalization
scheme  of the Gaussian approach 
as carried out, for example, in \cite{KERMANVAUTHERIN}
for the $\lap$ model is used as starting point
for  further investigation. 
It is  proposed the extremization of the renormalized
energy density with relation to the renormalized parameters
(coupling constant and mass).
Besides that some variables are placed in the complex
plane
to search   suitable (physical) values and eventual conditions 
for these parameters. 
The work is organized as follows. 
In the next section the Gaussian approximation is summarized:
the GAP equation (transcendental) 
is derived, obtained from the  regularized theory 
(with a cutoff).
The  renormalization procedure of the mass 
and coupling constant as proposed in 
 \cite{KERMANVAUTHERIN} is considered.
In sections 3, 4 and 5 values of the renormalized
mass, condensate and coupling constant which extremize the 
energy density are searched and analysed. 
They also could yield privileged values of the parameters 
with which the 
approximation may be more appropriated.
In some cases  instabilities are found for values of the parameters.
In section 6 a mathematical trick is used to search
 non transcendetal solutions or/and an expression which 
 constrains further the parameters. 
This is done by allowing some parameters to
be complex such that the imaginary part must in fact disappear
in the end of the calculation.
In the last section the results are summarized.

\section{ Gaussian approximation for the  $\lambda \phi^4$  model } 

The Lagrangian density 
 for the  scalar field   $\phi (\bx)$
 with bare mass   $m_0^2$
and coupling constant  $\lambda$ 
is given by:
\be \label{1}
\displaystyle{
{\cal L}(\bx) = \frac{1}{2}\left\{
\partial_{\mu} \phi(\bx) \partial^{\mu} \phi(\bx) - m_0^2 \phi^2(\bx) -
\frac{\lambda}{12} \phi^4(\bx) \right\}   }
\ee
The theory is quantized in the Schrodinger picture \cite{SCHRODINGER}
being the action of the field and momentum operators
over a state  $|\Psi[\phi]>$ given respectively 
by:
\be \ba{ll} \label{quant}
\displaystyle{ \hat{\phi} |\Psi> = \phi |\Psi> \;\;\;\;\;\;\;
\hat{\pi} = - i \hbar \frac{\delta}{\delta \phi} |\Psi >
}
\ea
\ee

In the static Gaussian approximation at zero temperature
the trial ground state wave functional $\Psi$ is parametrized by
the Gaussian:    
\be \label{4} \ba{ll}
\displaystyle{
\Psi\left[\phi(\vx )\right] = N  exp \left\{ -\frac{1}{4} \int
d \bx d \by \delta\phi(\vx ) G^{-1}(\vx ,\vy) 
 \delta \phi(\vy)
\right\}  , }
\ea
\ee
Where 
$ \delta\phi(\vx) = \phi(\vx)-\bar\phi(\vx) $ is the field 
shifted by the condensate, the point
 where the wave function is centered; 
the normalization factor is $N$, 
the variational parameters are   
the (classical) expected value of the field,
 $\bar \phi (\bx ) = < \Psi | \phi | \Psi >$, and the
quantum fluctuations represented by the two point function, i.e.,
the width  of the Gaussian:
$ G(\vx,\vy) = <\Psi |\phi(\bx) \phi(\vy)  | \Psi>$. 
In variational calculations  the averaged energy calculated with 
$\Psi [ \phi(\bx ) ]$
is to be  minimized to obtain the GAP equations.
In principle it would yield a maximum bound for the ground 
state (averaged) energy, although ultraviolet divergences make
this  not necessarily reliable. 
The minimization of the renormalized theory may be useful
for  this theoretical bound of the variational principle.
Each of these variational parameters represents one
component of the scalar field: the expected value in the 
ground state ("classical" part) and the two-point Green's
function with the mass of the quantum
which is decomposed into creation and annihilation operators
\cite{THESE}.

The average value of the Hamiltonian 
 is calculated and expressed in terms of the variational
parameters by means of expressions (\ref{quant}) and (\ref{4}).
It is given by:
\be \label{11a} \ba{ll}
{\cal H} & = \frac{1}{2} \left[ \frac{1}{4} G^{-1}(\bx,\bx) 
             - \Delta G(\bx,\bx) + m^2_0 G(\bx,\bx) + \frac{\lambda}{4} 
              G^2(\bx,\bx) + \right. \\
   & \left. + m^2_0 \bphi^2(\bx) + (\nabla \bphi(\bx))^2 
     + \frac{\lambda}{12}\bphi^4(\bx)
      + \frac{\lambda}{2} \bphi^2(\bx) G(\bx,\bx) \right].
\ea
\ee
Although in this expression the variational parameters were allowed to 
have spatial dependence they will be assumed to be constant.
Variations of the averaged energy density 
with respect to the variational parameters 
 yield the following GAP and condensate equations which define the 
ground state of the model:
\be \label{7} \ba{ll}
\displaystyle{ \frac{\delta {\cal H}}{\delta G(\bx,\by)} \rightarrow
0 =  -
\frac{1}{8}G^{-2}(\vx,\vy)   +
\frac{\Gamma(\vx,\vy)}{2} +
\frac{ \lambda}{2} \bar \phi(\vx)^2  \;\;\;\;\; (i) } \\
\displaystyle{
\frac{\delta {\cal H}}{\delta \bar{\phi}(\vx) } \rightarrow
0 = \Gamma(\vx,\vy)\bar \phi(\vy) +
\frac{ \lambda}{6}\bar \phi^2(\vx), \;\;\;\;\; (ii) }
\ea
\ee
Where
$\Gamma (\vx,\vy) = -\Delta  + \left( m_0^2 + 
\frac{\lambda}{2}  G(\vx ,\vx )
\right)\delta (\vx-\vy)$.
The Green's function $G$ may be written from expressions above as:
\be  \label{10}
G_0({\bx},{\by })=<{\bx}|\frac{1}{\sqrt{-\Delta + m^2}}|{\by}>
\ee
where  $m^2$  is given by the self consistent (transcendental)
GAP equation
(expression (\ref{7})):
\be \label{11} \ba{ll}
\displaystyle{ m^2 = m^2_0 + \frac{\lambda}{2} Trace G(x,x,m^2) 
+ \frac{\lambda}{2} \bphi^2 .}
\ea
\ee
An analogous expression holds for the case in which $\bphi = 0$, i.e.,
$$ \mu^2 = m^2 (\bphi=0) = m^2_0 + \frac{\lambda}{2} Trace G(x,x,\mu^2).$$
Expression (\ref{10}) is equivalent to the Feynman Green's function
with time integrated and with sign changed in the
imaginary part by replacing the self consistent mass by the
bare mass $m_0^2$.
The physical masses in the different phases may assume different values
from each other.
The condition of minimum for this procedure and its stability 
was partially investigated
in \cite{KMV} and it corresponds to analysing the second order variation
of the energy density with respect to the variational parameters.

From the above expressions it is seen that
the non zero solutions for the condensate, $\bphi$, can be written as:
\be \label{cond7} \ba{ll}
\displaystyle{ \bar{\phi}^2  = - 6 \frac{m_0^2}{\lambda}  
- 3 G (m^2) = 
\frac{ 3 m^2}{\lambda}.}
\ea
\ee
For $G=0$ the tree level value for $\bphi$ is obtained in terms 
of the bare mass.

The above expression for the Gaussian width (\ref{10}) (and its
inverse $G^{-1}_0$) can be calculated
in the momentum space with a regulator $\Lambda$ (cutoff) yielding
(for $\Lambda >> m^2$):
\be \ba{ll}
\displaystyle{ G(m^2) = \frac{1}{8\pi^2} \left( \Lambda^2 - 
m^2
Ln \left(\frac{ 2 \Lambda}{\sqrt{e} m} \right) \right), }\\
\displaystyle{ G^{-1}(m^2) 
= \frac{1}{8\pi^2} \left( 2 \Lambda^4
+ 2 m^2 \Lambda^2 - \frac{m^4}{4} - m^4 Ln \left(
\frac{2 \Lambda}{\sqrt{e} m} \right) \right),
}
\ea
\ee
where $d= 2/\sqrt{e}$. 
In the (local) 
limit of infinite cutoff the average energy 
and observables diverge and the divergences must be eliminated.
The renormalization procedure has been performed 
in  three dimensions for example in 
\cite{KERMANVAUTHERIN,STEVENSON,BRANCHINA,KMV}.

\subsection{ Renormalized parameters }

The renormalization procedure of the parameters of the
model is done as follows.
The  energy density of the 
symmetric phase, 
as well as its GAP equation (\ref{11}), is subtracted
from the corresponding  expression of the asymmetric
phase.
The GAP 
equation as defined in expression (\ref{11}) can be rewritten as:
\be \ba{ll} \label{MU2REN}
\displaystyle{ \mu^2 = m^2 + g_R \left( \bphi^2
+ \frac{m^2}{8 \pi^2} Ln \left( \frac{m}{\mu} 
\right) \right),}
\ea
\ee 
where the renormalized  parameters were defined as:
\be \label{flow} \ba{ll}
\displaystyle{ \mu^2 = m_R^2  \equiv \frac{ m^2_0 + 
\frac{\lambda \Lambda^2}{16 \pi^2} }{
1 + \frac{\lambda}{16 \pi^2} log\left( \frac{d \Lambda}{
\mu} 
\right)},
}\\
\displaystyle{ g_R = \frac{ - 
\frac{\lambda}{2} }{
1 + \frac{\lambda}{16 \pi^2} 
log\left( \frac{d \Lambda}{
\mu} 
\right)}.
}
\ea
\ee
In the first of these expressions 
 $m_R^2 \equiv \mu^2$ was chosen to produce the usual effective
potential  \cite{STEVENSON,KERMANVAUTHERIN}.
It is seen from the second of these expressions that 
in the limit of $\Lambda \to \infty$ the bare
coupling constant would go to zero in order to keep
$g_R$ finite if $\mu$ is kept constant. 
This is the ``triviality´´ problem.

The resulting subtracted energy density,
${\cH}_{sub} = {\cal H}(\bphi)
 - {\cal H}(\bphi = 0)$,  
is re-written in terms of the renormalized
mass, coupling constant 
and the mass scale 
 eliminating the cutoff.
It is given by:
\be \label{HSUB} \ba{ll} 
\displaystyle{ {\cH}_{sub} = \frac{m^2}{2} \bphi^2 + 
\frac{1}{4 g_R} \left( m^2 - \mu^2 \right)^2 + 
\frac{1}{128 \pi^2} \left( m^4 Ln \left(
\frac{m^4}{\mu^4} \right) - m^4 + \mu^4
\right) . }
\ea
\ee
The mass scale $\mu^2$ is not a free parameter in fact, it 
can be considered to be a function of the mass $m^2$ and
the coupling $g_R$ by the
GAP expression (\ref{MU2REN}).
Other approaches may be of interest for investigating
the variational method in the Schrodinger picture
\cite{STEFANOVICH}.
In the ground state the parameters $\bphi, m^2, \mu^2$ 
(for a given $g_R$)
are related by the GAP and condensate expressions shown above.
Any deviation of the respective numerical values from the ones related by
these expressions induce temporal evolution \cite{THESE}.

It is possible to verify whether the renormalized
GAP equation obtained from the regularized energy density
given by expression (\ref{MU2REN}) still is a GAP equation for
the energy density given by  expression  (\ref{HSUB}) in two
ways. 
The minimization of expression  (\ref{HSUB}) with relation 
to $m^2$ is done in the 
next section. 
However the integration of the GAP equation 
with relation to $m_2$ should result in an expression 
equal to  (\ref{HSUB}) if the order of performing 
renormalization and extracting the ground state 
does not change results.
The integral of the GAP equation is given by:
\be \ba{ll} \label{intgap}
\displaystyle{ \int \left(
- \mu^2 + m^2 + g_R \left( \bphi^2
+ \frac{m^2}{8 \pi^2} Ln \left( \frac{m}{\mu} 
\right) \right) \right)  d m^2 = } \\ 
\displaystyle{ = - \mu^2 m^2 + \frac{m^4}{2} + g_R  \bphi^2 m^2
+ \frac{g_R}{16 \pi^2} \left( \frac{m^4}{2} 
Ln \left( \frac{m^2}{\mu^2} \right) - 
\frac{m^6}{6 \mu^2} \right)  + 
C (\bphi, \mu^2) 
,}
\ea
\ee 
where $C(\bphi, \mu^2)$ does not depend on $m^2$.
This expression contains terms very different from the
renormalized expression (\ref{HSUB}).
This means either that the minimization of the 
regularized energy is not equivalent to the 
minimization of the renormalized one or/and that
the renormalization procedure has to be improved
to make both procedures coincident - if this 
is possible or  desirable.
This will be discussed below with the minimization
of the energy density with respect to $\bphi$.

\section{Energy density and  renormalized mass}

In this section 
the renormalized energy density ${\cH}_{sub}$ is extremized
with relation to the renormalized (physical) mass:
\be \ba{ll} \label{Hmmin}
\displaystyle{
 \frac{\partial {\cal H}_{sub}}{\partial m} = 0.
}
\ea
\ee
The roots of the resulting expression were calculated
by considering that $\bphi^2$ is in fact dependent on $m^2_R$ by
expression (\ref{MU2REN}).
The  resulting expression is given by:
\be \label{solutionmr} \ba{ll}
\displaystyle{ 0 = m^3 \left[ Ln^2
\left(\frac{m}{\mu}\right) a_1 + 
Ln \left(\frac{m}{\mu}\right) a_2 + a_3 
\right]
,}
\ea
\ee
where $a_i$ can be given in terms of 
$$J = 1 - \frac{g_R}{(8\pi)^2} = 1 - G_R, $$
 by:
\be \ba{ll}
\displaystyle{ a_1 = \frac{1}{g_R} J^2 
 + \frac{1}{32 \pi^2} 
,}\\
\displaystyle{ a_2 = \frac{2}{g_R} \left( -1 + J + 
\frac{J^2}{(32 \pi^2)} \right)+ \frac{1}{128 \pi^2}
\left( 1 + \frac{2 J^2}{(8 \pi)^2} \right)
,}\\
\displaystyle{ a_3 = \frac{1}{32 \pi^2} 
\left( 1+ \frac{g_R}{32 \pi^2} \right).
}
\ea
\ee
Expression (\ref{solutionmr}) is not equal to the GAP (\ref{solutionmr})
obtained from the minimization of the regularized energy density
with relation to $G(m^2)$.
There are therefore 
five solutions for the renormalized mass $m^2$ which can be written
in the following form:
\be \ba{ll} \label{mpm}
\displaystyle{ m^3 = 0, }\\
\displaystyle{ m^{\pm} = \mu \; exp( H^{\pm} ) ,}
\ea
\ee
where:
\be \ba{ll}
\displaystyle{ H^{\pm} = \frac{-a_2 \pm \sqrt{ a_2^2 - 4 a_1 a_0}
}{2 a_1}
.}
\ea
\ee
These solutions for $m^{\pm}$ can be viewed as having corrections
for the value of $\mu$ due to the self interaction through the 
parameters $H^{\pm}$ due to the appearance of $\bphi \neq 0$.
It is noted that there is a sort of ``energy scale'' invariance
 in these
expressions for $m^{\pm}$ with simultaneous changes in 
the mass renormalization parameter $\mu$.

The particular case of $m^2 = \mu^2$, for which the 
GAP equation is not necessarily valid because $\bphi \to 0$, 
is found for
\be \label{a0to0} \ba{ll} 
\displaystyle{ a_0 = 0, \;\;\; \to \;\;\; g_R = -32 \pi^2 .}
\ea \ee
This point may correspond to a restoration of the symmetry.

The  zero mass solutions correspond to a 
saddle point, they are not minima neither maxima of the energy
density.
If  the others solutions are minima  is checked via the 
positiveness of the second derivative:
\be \ba{ll}
\displaystyle{\frac{\partial^2 \cH_{sub}}{\partial m^2} = 
\frac{m^2}{(8\pi)^2} \left( 2 Ln
\left(\frac{m}{\mu}\right) a_1 + a_2 \right) > 0.
}
\ea
\ee
For the derivation of these expressions  the 
complete self consistency of the Gaussian equations was not considered. 
There has been used a truncation on the dependence on $\mu$, i.e.,
the dependence of $Ln(\mu/m)$ on $\mu$ (self consistency)
was considered only for $\mu$ not very different from $m$, i.e.
$\mu^2 = m^2 + \delta$ where $\delta << m^2$. Out of this range
the above solutions are not expected to be valid.

In Figures 1a and 1b the solutions of the above equations
($m^{\pm}/\mu$ from (\ref{mpm})) 
are shown as a function of $G_R = g_R/(32 \pi^2)$.
All the solutions of figure 1a, for $m^+$, correspond to stable solutions 
($d^2 \cH/ d m^2 >0$). The solutions of figure 1b, for $m^-$, are stable
for $G_R$  nearly equal or smaller than $-1.45$ or  
equal or greater  than nearly $1.25$.
The point $g_R=0$ is not plotted.
Values between $ -1 < G_R < 0$ do not correspond to physical 
stable values of the condensate
as it will be shown below, in expression (\ref{COUPLING1}).
In the limits of $g_R \to \pm \infty$ we obtain analytically 
that 
either $ m = \mu$ or $m=0$.
For the case $\mu \to \infty$ the renormalized coupling constant
$G_R \to 0$.
While the solution $m^{-}_R$ in the weak coupling regime 
can be identified to the renormalization point usually considered
(for $\mu >> m$ and/or the cutoff going to infinite)
there is another stable solution $m^+$ for which $\mu \simeq m^+$.

\section{ The condensate: $\bphi$}

The variational equation for the condensate (expression (\ref{7} (ii))) 
is obtained from the 
regularized energy density $\cH_{reg}$. The minimization of 
the renormalized energy density is done in the following:
\be \ba{ll}
\displaystyle{ \frac{\partial {\cal H}_{sub}}{ \partial \bphi} = 0
.}
\ea
\ee
For this derivation the GAP equation provides the
dependence of the mass on the condensate, i.e., 
$m^2 (\bphi)$ and $\mu^2 \equiv m^2 (\bphi =0)$ 
is kept constant. 
It yields the following expressions:
\be \label{phi0} \ba{ll}
\displaystyle{ \bphi = 0,}\\
\displaystyle{ \bphi^2 = - \frac{m^2}{g_R} 
\left( 1 + \frac{1}{8 \pi^2} Ln \left(\frac{m}{\mu}\right) 
\right) .}
\ea
\ee
This last expression may coincide with the expression of $\bphi_0$
obtained from the minimization of the regularized energy density
(expression (\ref{cond7}))
depending on the relation between 
$\lambda$ and mass scale $\mu$ as it will be shown below.
However it is not completely 
consistent with the GAP equation (\ref{MU2REN})
which is obtained from the minimization of the reguralized
energy density with respect to the mass $m$ in the asymmetric
phase and then renormalized.
To make these expressions compatible it would be necessary
to consider the following alternatives  for these expressions:
\be \ba{ll} 
\displaystyle{ 
\mu^2 \neq m_R^2, \;\;\;\;\;\;\;\;
\mbox{or} \;\;\;\;\;\;\;
\mu^2 = (g_R - 1) \frac{m^2}{8 \pi^2} Ln \left( 
\frac{m}{\mu} \right), }
\ea \ee
where $m_R^2$ is the one of expression (\ref{flow}).
It is not clear whether these identifications are
 reasonable or if they imply  a meaningful
loss of generality.
The limit of $g_R = 1$ does not seem to be reasonable.
Therefore the two minimization procedures (of the regularized
and the renormalized energy densities with respect to
the regularized and renormalized parameters respectively)
do not seem to yield necessarily the same expressions 
for the parameters in the ground state.
Nevertheless it is worth to remember 
that renormalization is performed basically
from the regularized GAP equation.

From the  expression (\ref{phi0}) 
the following conditions 
to obtain real values of $\bphi$ can be considered:
\be \label{COUPLING1} \ba{ll}
\displaystyle{ if:  g_R > 0  \to 
Ln \left(\frac{m}{\mu}\right) < - 8 \pi^2 ,}\\
\displaystyle{ if:  g_R < 0  \to Ln 
\left(\frac{m}{\mu}\right) > - 8 \pi^2 .}
\ea
\ee

The energy density is expected be stable for 
the condensate values found in expression (\ref{phi0}).
This minimum is verified by calculating the second derivative
of the energy density with relation to $\bphi$, i.e.:
$\partial^2 \cH_{sub} /\partial \bphi^2 > 0$.
Its positiveness  corresponds to the condition:
\be \label{COUPLPOS} \ba{ll}
\displaystyle{ g_R \left( 1 + \frac{g_R}{32 \pi^2} 
\right) > 0 .}
\ea
\ee
From this it is seen  that for positive coupling constant
$g_R$, it can assume any value (from this stability criterium)
whereas if $g_R < 0$ one would have to consider $g_R < -32 \pi^2$.
Expressions (\ref{COUPLING1}) and (\ref{COUPLPOS}) may
 correspond to constraints for the 
values that the renormalized coupling may assume in order to 
yield stable real ground states.

Expression (\ref{phi0}) can be written as:
\be \label{bphinovo} \ba{ll}
\displaystyle{ g_R \bphi^2 = - m^2 \left( 1 + \frac{1}{8 \pi^2}
Ln \left( \frac{m}{\mu} \right) \right).
}
\ea
\ee
When $\mu = m \; exp(8 \pi^2)$ it follows that
 either $\bphi=0$ or $g_R=0$
in the asymmetric phase of the potential. 
This may correspond to the so called symmetry restoration 
when the condensate disappears at a particularly
high excitation energy, i.e., the symmetry is restored.
A different solution for the particular limit of $\bphi = 0$
was found in expression (\ref{a0to0}) where the energy density
is minimum with relation to the mass for $m^2 = \mu^2$.

The above expression for the condensate 
(\ref{phi0}) can  be equated to the 
previous (regularized) one  (\ref{cond7}).
Taking into account the expression of the renormalized coupling constant
in terms of the bare one (expression (\ref{flow}))
this  can be written as:
\be \label{constraint} \ba{ll}
\displaystyle{ \lambda = 
\frac{16 \pi^2}{Ln \left(\frac{\Lambda d}{\mu}\right) }
\left( -1 + \frac{3}{2 \left( 1 + \frac{1}{8\pi^2} Ln \left(
\frac{m}{\mu} \right) \right) } \right) .
}
\ea
\ee
If the cutoff is sent to infinite the bare coupling constant
assumes different values depending on the ratio of $\mu/m$.
For example, there is  a case in which $\lambda = 0$
if either $\Lambda \to \infty$ for finite $\mu$
or:
\be \ba{ll} \label{masses-2}
\displaystyle{ \frac{m}{\mu} = exp (4\pi^2),}
\ea \ee
being therefore $m^2 >> \mu^2$. 
Varying $\mu$ together with $\Lambda$
there may have non zero  $\lambda$ solutions. 
For $\Lambda/\mu$ finite, the coupling $\lambda$
may even diverge when: 
\be \ba{ll} \label{masses-3}
\displaystyle{ \frac{m}{\mu} = exp (-8 \pi^2).}
\ea \ee
This is the same point found above (for expression (\ref{bphinovo}) 
for the possible restoration
of the symmetry.

It is worth emphasizing that
it has been assumed, as it usually is, that the minimum of the 
effective potential with relation to the condensate coincides
necessarily with its minimum in respect to the physical mass $m_R^2$
in the regularized theory.

\section{ Analysis of the renormalized 
coupling constant }

Analogously to what was done for the renormalized mass in the 
preceeding section  the extremization of the 
renormalized energy density with respect to
the renormalized coupling constant is done in this section. 
Moreover one relevant subject for any approximation method is 
the understanding of the range of values of the parameters of the 
model (as mass and mainly coupling constants) 
for which the approximation is more appropriated. 
The extremization is found from:
$$ \frac{\partial {\cal H}_{sub}}{ \partial g_R} = 0.$$
It is  considered, in the following, a truncation
of the self consistency of the GAP equations.
This is done 
 by taking the scale parameter to be close
to the  mass $\mu^2 = m^2 + \delta$, where $\delta << m^2$ 
is determined from the GAP equation self consistently.
From the renormalized GAP equation (expression (\ref{MU2REN})) 
it follows that:
\be \label{delta} \ba{ll}
\displaystyle{ \delta = \frac{g_R \bphi^2}{ 1 + \frac{g_R}{16 \pi^2}}
.}
\ea
\ee
Since $\delta << m^2$ either $\bphi$ is large or $g_R$ is very large
for positive coupling $g_R$.
The minimization of the renormalized energy yields the following 
third order algebraic expression:
\be \label{MINICOUPL} \ba{ll}
\displaystyle{ (G_R')^3 + (G_R')^2 \left( 3 + (1 + H)\frac{1}{16 \pi^2} 
\right) + G_R' \left( 3 + (1 + H)\frac{3}{2} + 
(1+ H)\frac{1}{32 \pi^2} \right) + 1 + \frac{(1+H)^2}{2} = 0,
}
\ea
\ee
where 
$$H = \frac{Ln \left(\frac{m}{\mu}\right)}{(8 \pi^2)} , \;\;\;\;\;\;
G_R' = \frac{g_R}{(16 \pi^2)}.$$

In figures 2a, 2b and 2c the solutions of expression (\ref{MINICOUPL})
are showed as  function of a limited range of 
$H$, i.e.,  $Ln(m/\mu)$. 
Figures 2b and 2c exhibit the same behavior.
It is plotted only the region in which the above truncation scheme
of the self consistency may be  expected 
to be reliable.
The values for $g_R$ are large and obtained for $m^2 \sim \mu^2$ 
which 
cannot be simultaneously compatible with the results of 
 Figures 1a and 1b (from the minimization of the energy 
with relation to the mass $m^2$) - in this limit of $\delta << m^2$. 
The point $H = -0.001$ corresponds to $m/\mu = 0.985$ which is not
obtained for the larger values of $g_R$ from figures 1a and 1b.

These solutions
(\ref{MINICOUPL}) may nevertheless correspond to 
 minima or maxima of the solutions of the above equation is
also  verified. For this the 
positiveness of the second derivative is calculated:
$\frac{\partial^2 {\cal H}_{sub}}{\partial g_R^2} > 0$.
All the solutions 
have a negative second derivative corresponding to maxima, instead
of minima, of $\cH_{sub} (g_R)$.

The solutions for the 
coupling constant of expression (\ref{MINICOUPL}) 
depend only on the ratio $m/\mu$ and not 
on the absolute values of these parameters. 
This may be seen as  a sort of 
energy scale invariance for different 
physical processes (eventually in different systems) at different
energy scales with different physical masses.

\subsection{ Fixing the energy density}

For the analysis of the system
 with an energy density  given by $\cH_{sub}$ and
a given mass scale, $\mu$, (acceptable)
values for the renormalized coupling constant and mass may be suggested
such that the values remain in the physically 
allowed part of the phase space of the model \cite{FLB2001PRD}.
This corresponds to fix the renormalized mass and 
energy density  (${\cal H}_{sub}$) and to calculate
the resulting physical coupling constants for the process involved
at a scale $\mu$.
A third degree algebraic equation is obtained and which can
be written as:
\be \label{CONDCOUPL} \ba{ll}
\displaystyle{ g^3_R \frac{H^2}{128 \pi^2} +
g_R^2 \left( \frac{H^2}{4} - \frac{2 H^2}{x} \right)
+ g_R \left( -\frac{H(H+1)}{2} + \frac{1}{x} \left(
\frac{H}{4} - 1 + H^2 \right) - \frac{{\cal H}_{sub}}{m^4}
\right) -\frac{1}{4} + \frac{H^2}{4} = 0.
}
\ea
\ee
where $x= 128 \pi^2$ and $H=Ln(m/\mu)/(8\pi^2)$.
This expression also presents a sort of  ``energy scale'' invariance for 
the parameters $m/\mu$ unless for the term which depends on the
total energy density, if ${\cal H}_{sub}$ scales differently from $m^4$.

In figures 3a, 3b and 3c the solutions of this algebraic
equation are shown as functions of $H$ for a fixed energy density 
$\cH = (100 MeV)^{-4}$ and $m_R = 100 MeV$. 
The coupling $g_R$ can be strong in the region of 
$\mu \simeq m$.
In particular in the limit of $\mu = m$ an unique
value is found, it is given by:
\be \label{grHfix} \ba{ll} 
\displaystyle{ g_R = - \frac{1}{4 \left( \frac{\cH^{sub}}{m^4} 
+ \frac{1}{128\pi^2} \right)
}.
}
\ea
\ee
For $H \to - \infty$, which is 
equivalent to $\mu/m \to \infty$, it follows 
$g_R \to 0$ as seen in figure 3a. 
However
solutions of figures 3b and 3c do not correspond to $g_R \to 0$, but
to a finite (quite strong) value close to 10.

\section{
The transcendental character of the GAP equation} 

In this section an heuristic  trick is used to extract  analytical
non transcendental solutions from the  GAP equation or to provide
possible further relations among the parameters.
Firstly it is considered that the  the 
mass scale and renormalized mass  may have  imaginary parts:
\be \ba{ll} \label{mass-comp}
\displaystyle{ \mu^2 \to \nu^2 = r e^{i \theta} , 
 \;\;\;\;\;\;\;  m^2 \to \tau = t e^{i \omega} ,
}
\ea \ee
where $r, s , \theta, \omega$ are respectively modulus and phases.
With these parametrizations the GAP equation (\ref{MU2REN}) 
can be written as:
\be \ba{ll} \label{GAPcomp1}
\displaystyle{ \left(r cos \theta - t cos \omega - g_R \bphi^2
- D (t \, cos \omega \, Ln (t/r) - (\omega - \theta) t sin \omega  )\right) + 
} \\
\displaystyle{
+ i \left[ t \; sin \omega - r \; sin \theta + D ( \omega t \; 
cos \omega - t \theta \; cos \omega + t \; sin \omega \, Ln (t/r) )
\right] = 0,
}
\ea 
\ee
where  $D = \frac{g_R}{8 \pi^2}$.
Both the real and the imaginary parts in this expression have 
basically the same structure of the usual GAP equation.
It is worth emphasizing that requiring the imaginary part to 
disappear in the GAP equation
 implies real mass parameters if only one of them is 
placed in the imaginary plane, i.e., either $\theta = 0$ or $\omega = 0$.
In these cases the parametrization in the complex plane is just a trick
to reduce the transcendental character of the GAP equation (\ref{MU2REN}).

Requiring the GAP equation (\ref{GAPcomp1}) to have only real component 
(this is considered to be a stable system) the imaginary
part is set to zero. The expression still is quite complicated but
the analysis of  some particular cases will be very useful.
For $\omega =0$ it follows that:
\be \ba{ll} \label{GAPcomp2}
\displaystyle{ r sin \theta = - D \, t \, \theta ,
}
\ea \ee
This  may be written as:
\be \ba{ll} \label{GAPcomp4}
\displaystyle{ cos \theta = \sqrt{ 1 - \frac{B^2 \theta^2}{r^2} }
.
}
\ea \ee
In this  case the self consistent character of the 
GAP equation remains
strong. The real part of expression (\ref{GAPcomp1}) 
keeps the same form of expression (\ref{MU2REN}) 
basically with the mass parameters $m^2, \mu^2$ replaced by $r, t$.

For $\theta = 0$ (and $\omega \neq 0$)
the resulting expressions for the real part of the 
GAP equation and its  imaginary part (to be equated to zero)
 can be obtained from expression (\ref{GAPcomp1}). 
They can be written as:
\be \ba{ll} \label{GAPcomp1a}
\displaystyle{ r - t \, cos \omega - g_R \bphi^2
- D (t \, cos \omega \, Ln (t/r) - \omega  t \, sin \omega  ) = 0
,} \\
\displaystyle{
+ i \left[ t \, sin \omega + D ( \omega t \,
cos \omega + t Ln (t/r) \; sin \omega )
\right] = 0.
}
\ea 
\ee
It does not provide simpler solutions  and therefore they are not shown.
The resulting number of free parameters is not reduced because although
there is one more expression ($\Im m (GAP)$)
there also is one extra variable ($\omega$).

Since the phases are auxiliar parameters it is reasonable to assume
they are very small without (great) loss of generality for the results.
The expression for the imaginary part of the GAP equation
in the limit when $sin (\theta) \sim \theta$ and 
$sin (\omega) \sim \omega$  is given by:
\be \ba{ll} \label{smalltw1}
\displaystyle{ 
\omega t \left( 1 + D + D Ln \left(\frac{t}{r}\right) \right)
= \theta \left( r + D t \right)
.}
\ea \ee
This expression can be regarded as fixing the ratio $\theta/\omega$.
Several particular cases  are analyzed below although 
the more interesting case is obtained for $\omega, \theta$ non zero
and very small.

{\bf (1)}
Assuming the phases are equal $\theta = \omega$ expression
(\ref{smalltw1}) reduces to:
\be \ba{ll} \label{tGAPcomp1}
\displaystyle{ r - t = 
 D \, t \, Ln \left(\frac{t}{r}\right)
,}
\ea \ee
which fixes the ratio $r/t$ or correspondently $m^2/\mu^2$.
This expression is only  consistent
with the renormalized GAP equation \ref{MU2REN} for $\bphi = 0$ 
(which is obtained from the
minimization of the regularized energy density).
Besides that it was mentioned above that, since $\omega \neq 0$
and $\theta \neq 0$, it is not clear whether $\mu^2$ and $m^2$
remain real although the GAP equation is necessarily real.
This may happen because in this case the imaginary part of both
parameters may cancel with each other to result a real GAP 
equation instead of allowing for independent cancelation.
On the other hand each angle ($\omega$ or $\theta$)
may be set to zero separatedly.

{\bf (2)} For $\omega = 0$ it follows:
\be \ba{ll} \label{tGAPcomp2}
\displaystyle{ r \simeq - D \, t,
}
\ea \ee
which also fixes the ratio $m^2/\mu^2$ being a real number  only for
$g_R < 0$.

{\bf (3)} 
For $\theta = 0$, expression (\ref{smalltw1}) is computed up to 
the order of $O(\omega^2)$ and it reduces to:
\be \ba{ll} \label{tGAPcomp3}
\displaystyle{ \omega^2 = 2 D + 2 + 2 Ln \left( \frac{t}{r} \right)
,}
\ea \ee
where it has been assumed that $sin \omega \sim \omega$.
In this case it is reasonable to consider $\omega^2 \sim 0$ leading 
to the expression:
\be \ba{ll} \label{tGAPcomp4}
\displaystyle{ \frac{t}{r} = exp (-2 -D ).}
\ea \ee

If $\omega \neq 0 $ it will appear in the real part of the GAP equation
and therefore the number of free parameters in the renormalized equation 
does not diminish with the new parametrization.
Therefore $\omega = 0 $ would be the only possibly interesting case.
This does not happens because of expression (\ref{tGAPcomp2}) which
imposes negative coupling $g_R < 0$.

The real part of the GAP equation for small angles
 keeps nearly the form of the
original GAP, it can be written as:
\be \ba{ll} \label{GAPcomp30}
\displaystyle{ t - r + g_R \, \bphi^2 + 
D \left[ t Ln \left(\frac{t}{r}\right)
 - t \, \omega \, (\omega - \theta) \right]
= 0 ,}
\ea \ee
where either $r$ or $t$ can be written as a function of the other 
by means of the constraints of the imaginary parts from  
the expressions (\ref{tGAPcomp1}), (\ref{tGAPcomp2}) or (\ref{tGAPcomp3}).
In this third case the auxiliar parameter $\omega$ was not eliminated
(although $\theta = 0$). 
However for very small phases the expression
(\ref{GAPcomp30}) reduces to the usual real GAP equation 
(\ref{MU2REN}). 
In this case the real part of the GAP equation is 
the same as expression (\ref{MU2REN}) written as:
\be \ba{ll} \label{GAPcomp50}
\displaystyle{ t - r + g_R \, \bphi^2 + 
D \, t \,  Ln \left( \frac{t}{r} \right)
= 0 .}
\ea \ee

Simultaneously the renormalized energy density must be a real number.
The imaginary part due to the introduction of parametrization 
(\ref{mass-comp}) has to disappear. 
However it is easy to notice from  expression (\ref{smalltw1}) 
that the resulting expression for the imaginary part of $\cH_{sub}$
will be quite complicated. 
Below it will be assumed that the phases have small values. 
This should not impose great 
limitations in the results because they are auxiliar parameters.
With this assumption several simplifications occurs because:
$sin (\theta) \sim \theta$ and $sin (\omega) \sim \omega$.
The result for the imaginary part of the energy density, up to 
first order in the phases, will
be given by:
\be \ba{ll} \label{IMH}
\displaystyle{ \Im m (\cH_{sub}) = \omega \left( 
\frac{ t \bphi^2}{2}
+ 2 t^2 A_- + \frac{t r}{2 g_R} + \frac{r^2 }{64 \pi^2}
\right)
+ \theta \left( 2 A_+ r^2 - \frac{r t}{2 g_R} + \frac{r^2}{32 \pi^2}
Ln \left( \frac{t}{r} \right) - r^2 
\right)
\to 0,
}
\ea
\ee
Where 
$$A_{\pm} =   \frac{1}{4 g_R} \pm \frac{1}{128 \pi^2}. $$
One of these variables ($A_+$) can be identified with
the solution of fixed ${\cal H}_{sub} (\omega=\theta=0)$ 
for $\mu = m$ given by expression (\ref{grHfix}):
\be \ba{ll}
\displaystyle{ A_+ = - \left. \frac{{\cal H}_{sub}}{m^4} 
\right|_{\mu = m}. }
\ea \ee
Expression (\ref{IMH}) still is very complicated
and it can also be used to fix the ratio $\theta/\omega$
which can be  equated to the same ratio obtained from 
expression (\ref{smalltw1}). 
However this has been written for ${\cal H}_{sub}$ in the form
written in (\ref{HSUB}), which can be written differently by 
means of the GAP equation for $m^2 = m^2(\mu^2)$.
This allows to write an equation of $r$ as a function of $t$ 
and eliminate one of these variables.
The resulting identity reads:
\be \ba{ll} \label{thetaomega}
\displaystyle{ 
\frac{\omega}{\theta} = - \frac{2 A_+ r^2 - \frac{r t}{2 g_R} 
+ \frac{r^2}{32 \pi^2}
Ln \left( \frac{t}{r} \right) - r^2 }{\frac{ t \bphi^2}{2}
+ 2 t^2 A_- + \frac{t r}{2 g_R} + \frac{r^2 }{64 \pi^2}
}
= - \frac{ \frac{r}{t} + D 
}{\left( 1 + D + D Ln \left(\frac{t}{r}\right) \right)
}
.}
\ea \ee
In this expression the same parameter is used: $D= g_R/(8 \pi^2)$.
This (highly non transcendental) expression appears in addition 
to the usual real part of the GAP equation, expression (\ref{GAPcomp50}),
making a system of two algebraic expressions with two variables
($r, t$). 
 $g_R$ is another  input/free
parameter.

\section{ Summary}

A further analysis of the usual 
renormalization scheme for the variational
Gaussian approximation was done. 
The renormalized energy density was extremized  with respect to 
the renormalized mass and coupling and to the condensate.
Concerning the extremization with respect to the mass, 
five solutions were found, two of which which can correspond to stable
vacua in specific ranges of the renormalized coupling constant.
For this it was considered that
the mass scale $\mu$ is close to the physical mass. 
A sort of  ``energy scale'' invariant  algebraic expression was found in
this calculation. In other words, changes in the renormalized
(physical) mass $m^2$ with corresponding change in the renormalization
mass scale parameter $\mu^2$ yield the same solutions.

Values for $\bphi$, in the vacuum, were also found
by minimizing the renormalized energy density with relation 
to it. The resulting expression is not completely 
consistent with 
the renormalized GAP equation unless the expression
(\ref{flow}) is modified such that $\mu^2 \neq m_R^2 \to 0$.
From this expression it was pointed out that either the 
``condensate'' or $g_R$ disappears when the mass scale 
(introduced in the renormalization procedure)
assumes the value 
$$\bphi \left( \mu = m \; exp(8 \pi^2) \right) = 0 .$$
This can be seen
as a restoration of the spontaneous  symmetry breaking.
With this value for $\mu$, 
the bare coupling $\lambda$
may also diverge for 
$\Lambda/\mu$ finite, as shown in expression (\ref{constraint}).

Particular values of the renormalized 
coupling constant which extremize the energy density 
were also found.
The coupling constant may constraint the values 
of the renormalized mass which yield maxima of the 
energy density (effective potential).
The extremization of the effective potential with respect to 
the coupling constant was also performed in the limiting 
case that the mass scale $\mu$ is close to the physical mass. 
This is a way 
of truncating the self-consistency of the approximation.
Another kind of ``energy scale'' invariant expression was also obtained.
No minima of the energy density with relation to 
$g_R$ was found (considered without the whole
renormalization group equations) within the 
truncation scheme which was adopted.
The renormalized energy was also fixed to provide
specific values for the coupling constant as a function
of the energy density and the mass, which disappears in the 
limit of $\mu = m$.

The masses were allowed to assume complex values
 to search non-transcendental
solutions for the GAP equation and other relations among the parameters
reducing the number of free parameters. 
The imaginary part is 
required to be zero at the end of the calculation 
producing another expression which relates 
the mass, coupling and the renormalization scale parameter.
This parametrization
for the imaginary part may lead to new relation between the parameters
 reducing the number of free variables.
 These imaginary parameters may be required to be very small 
($sin (\omega) \sim \omega$ or $sin (\theta) \sim \theta$).
The same parametrization is applied to the energy density which also
must be a real number. The number of free parameters ($m^2$ or $\mu^2$, 
and $g_R$) is reduced and non transcendental solutions may
result such as that of expression (\ref{tGAPcomp2}).

The ground state in the framework of the variational
approximation is found by the 
 minimization with respect to the two point function $G(\bx,\by, m^2)$
(which is a function of the physical mass $m^2$ or 
$\mu^2 = m^2 (\bphi=0)$)
and to the condensate $\bphi$ - they  are the
variational parameters (given in expressions (\ref{7})). 
Although they are regarded initially as independent variables,
the GAP equation (for ground state) relates them.
While the GAP equation is used for the renormalization of 
the bare parameters in the vacuum, 
 expression (\ref{phi0}) was calculated from 
renormalized expressions for the ground state. 
However there is nothing really defined about the behavior
of the renormalized parameters  in excited states.
 It was shown with sections  2.1, 3 and 4 that 
the minimizations of the renormalized energy with relation to 
the mass and $\bphi$ yield different ground state (GAP)
expressions from the ones obtained by the usual variational
procedure for the regularized theory.
This may have several meanings. 
It may not be evident whether these
variational parameters are 
really or completely suitable as  independent parameters
 for the 
Gaussian approximation and extensions 
(or leading order large N, Hartree Bogoliubov)
or even 
in the {\it exact ground state}, i.e., the energy must be minimum 
 with respect to  
particular combination(s) of these (or other) (physical?) variables.
Notwithstanding 
the minimization of the 
regularized energy may not be equivalent to the 
minimization of the renormalized one  because in the
regularized theory there still are other (bare) parameters
which are eliminated in the renormalization procedure 
corresponding to a sort of "hidden dependences" among them.
It may also be that
the renormalization procedure has to be improved such
as to make both ways of obtaining the ground state expressions
equivalent.
In this case the renormalization procedure would be allowed to be
done at any moment independently of the order of the 
the variation, renormalization and
extraction of observables within a certain nonperturbative approach.

\vspace{1cm}

{ \Large{ \bf Acknowledgements}  }

This work has been partially supported by FAPESP, Brazil. 
F.L.B. thanks A.J. da Silva, M. Gomes for brief discussion and
  O. \'Eboli,
J. Polonyi for reading an earlier version of the work.

\vspace{2cm}

{\bf Figure captions}

\vspace{1cm}

\noindent{ \bf Figure 1a -} First solution of expression (\ref{solutionmr})
 - a new GAP equation - for 
the ratio of the renormalized mass to the mass scale $\mu$  as a 
function of $g_R/(8\pi^2)$.

\vspace{1cm}

\noindent{ \bf  Figure 1b -} Second solution of expression (\ref{solutionmr}) 
- a new GAP equation - for 
the ratio of the renormalized mass to the mass scale $\mu$  as a 
function of $g_R/(8\pi^2)$.

\vspace{1cm}

\noindent{ \bf Figure 2a -}  First solution of expression (\ref{MINICOUPL}) 
for the renormalized coupling constant -
from the minimization of the energy density with relation to the renormalized 
coupling constant - as a function of $H = Ln(m_R/\mu)/(8\pi^2)$.

\vspace{1cm}

\noindent{ \bf  Figure 2b -} Second solution of expression (\ref{MINICOUPL}) 
for the renormalized coupling constant -
from the minimization of the energy density with relation to the renormalized 
coupling constant - as a function of $H = Ln(m_R/\mu)/(8\pi^2)$.

\vspace{1cm}

\noindent{ \bf Figure 2c -} Third solution of expression (\ref{MINICOUPL}) 
for the renormalized coupling constant -
from the minimization of the energy density with relation to the renormalized 
coupling constant - as a function of $H = Ln(m_R/\mu)/(8\pi^2)$.

\vspace{1cm}

\noindent{ \bf  Figure 3a -} First solution for the 
renormalized coupling constant of  expression (\ref{CONDCOUPL}) -
fixing $\cH = (100 MeV)^4$ and $m_R= 100 MeV$ - as a function of
$H = Ln(m_R/\mu)/(8\pi^2)$.

\vspace{1cm}

\noindent{ \bf Figure 3b -} Second solution for the 
renormalized coupling constant of  expression (\ref{CONDCOUPL}) -
fixing $\cH = (100 MeV)^4$ and $m_R= 100 MeV$ - as a function of
$H = Ln(m_R/\mu)/(8\pi^2)$. 

\vspace{1cm}

\noindent{ \bf Figure 3c -} Third solution for the 
renormalized coupling constant of  expression (\ref{CONDCOUPL}) -
fixing $\cH = (100 MeV)^4$ and $m_R= 100 MeV$ - as a function of
$H = Ln(m_R/\mu)/(8\pi^2)$. 

\newpage

\begin{figure}[htb]
\epsfig{figure=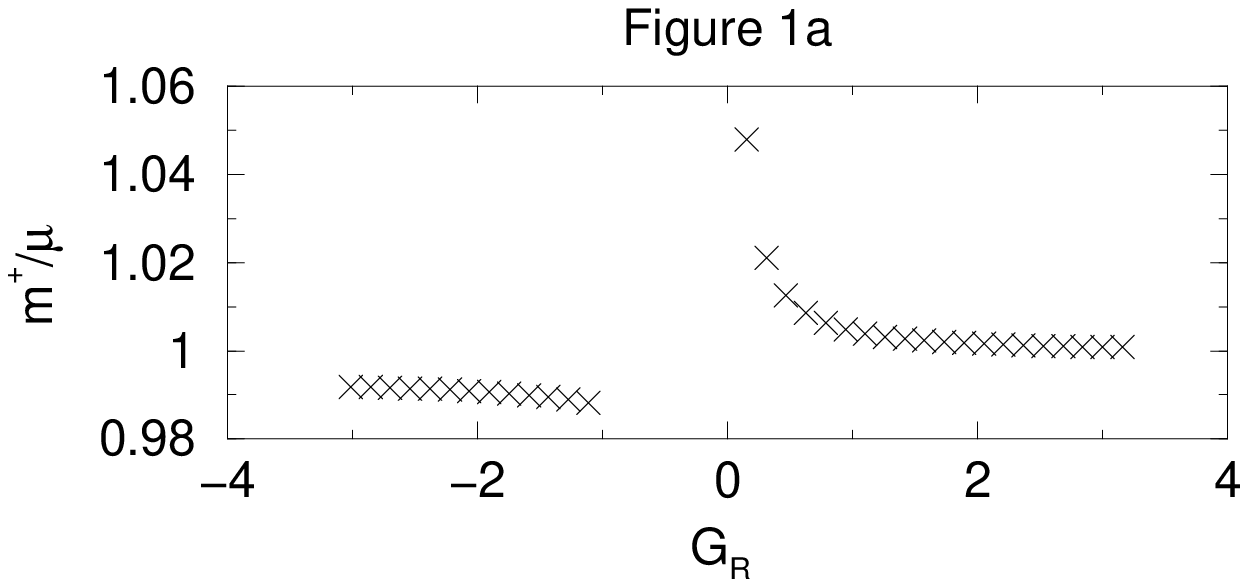,width=14cm} 
\end{figure}

\begin{figure}[htb]
\epsfig{figure=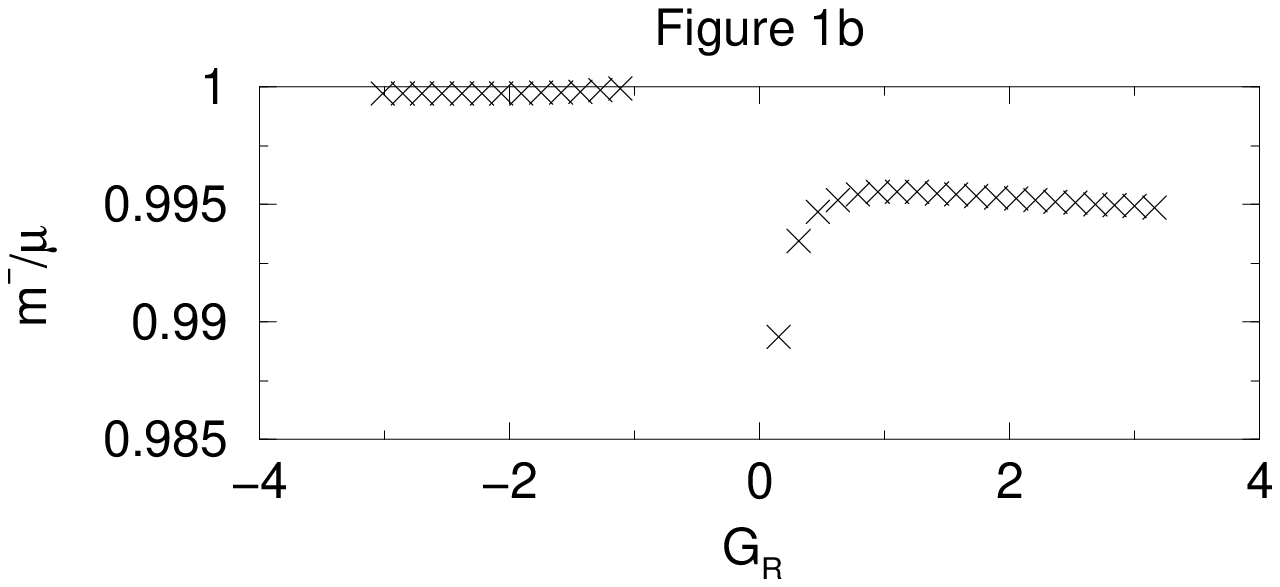,width=14cm} 
\end{figure}

\begin{figure}[htb]
\epsfig{figure=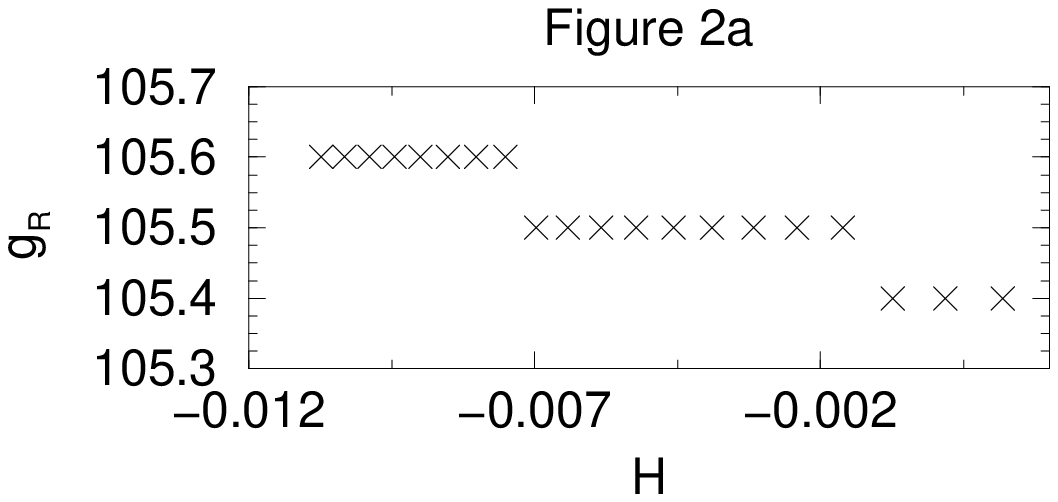,width=14cm} 
\end{figure}

\begin{figure}[htb]
\epsfig{figure=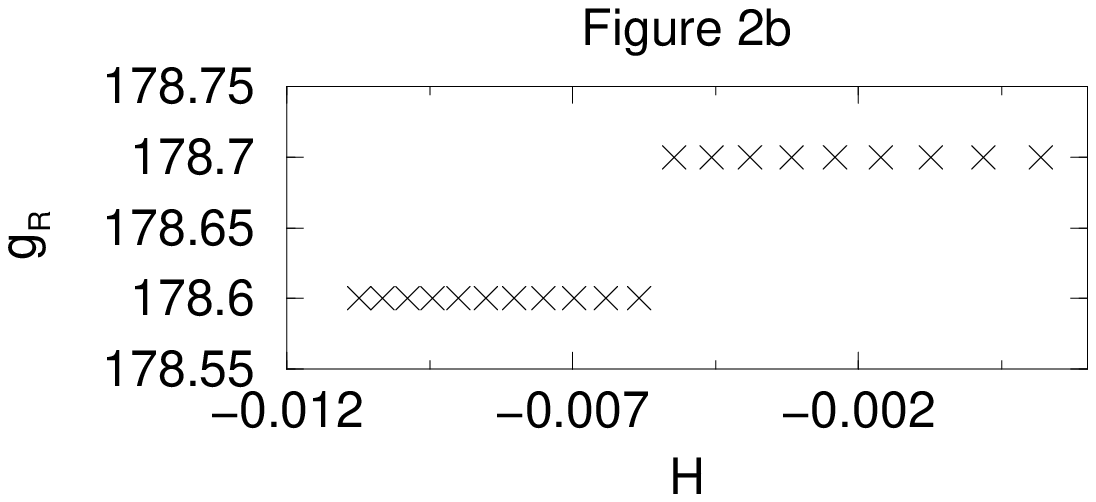,width=14cm} 
\end{figure}

\begin{figure}[htb]
\epsfig{figure=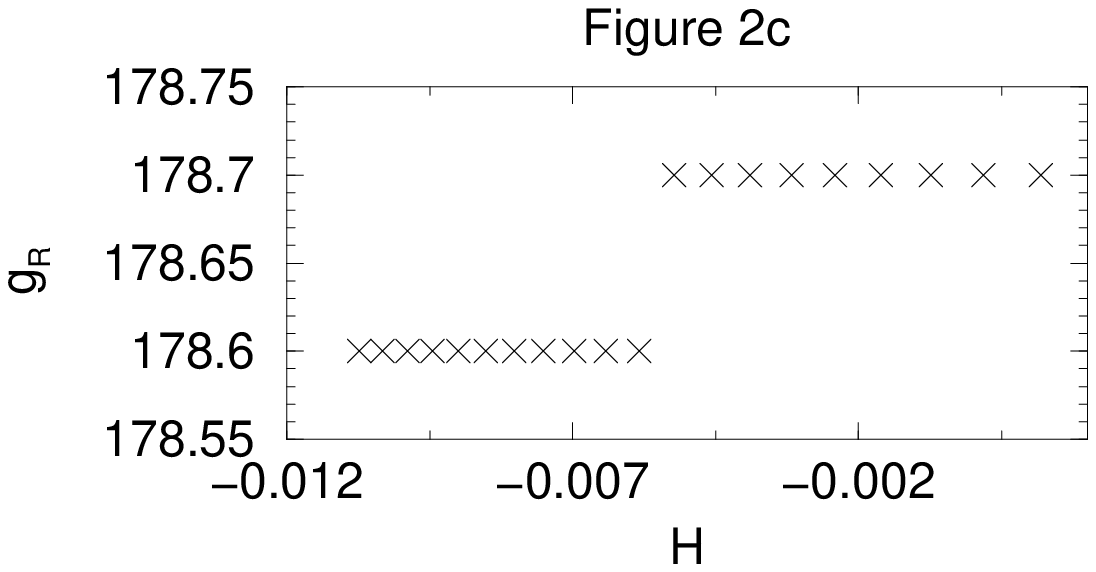,width=14cm} 
\end{figure}

\begin{figure}[htb]
\epsfig{figure=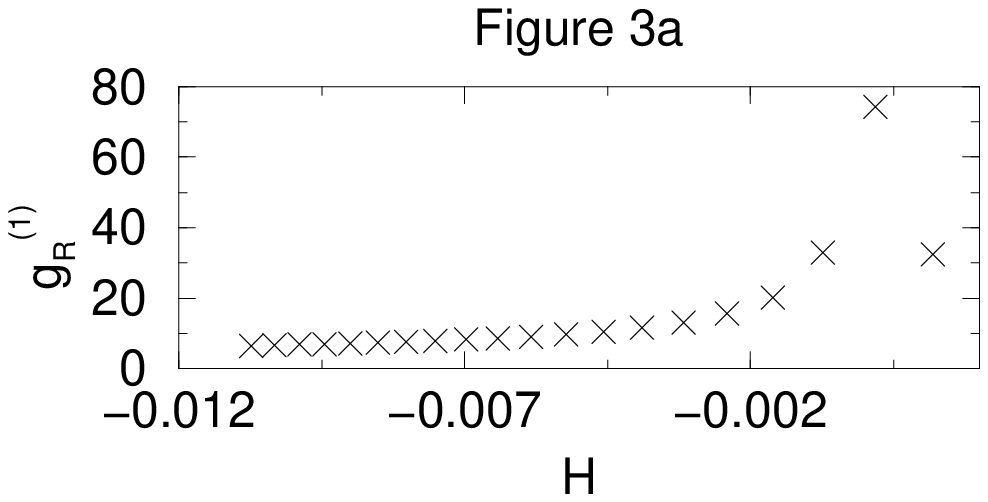,width=14cm} 
\end{figure}

\begin{figure}[htb]
\epsfig{figure=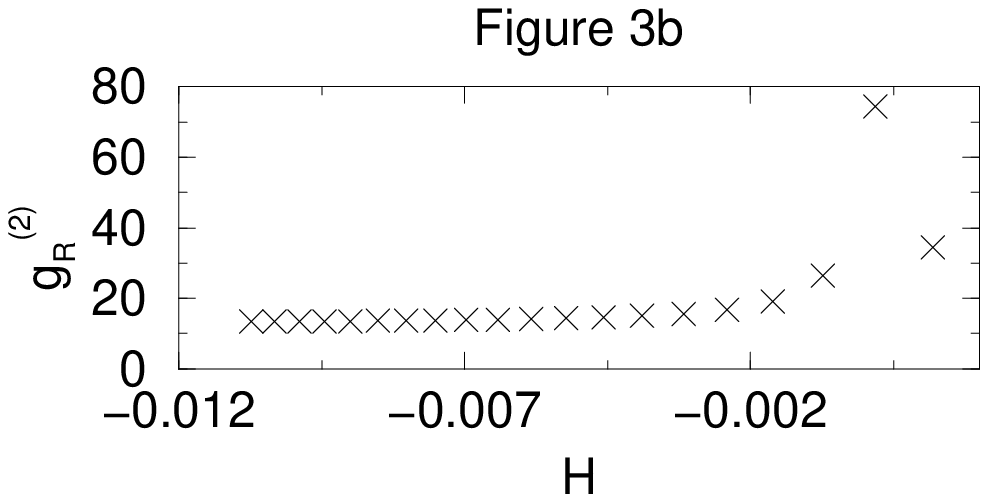,width=14cm} 
\end{figure}

\begin{figure}[htb]
\epsfig{figure=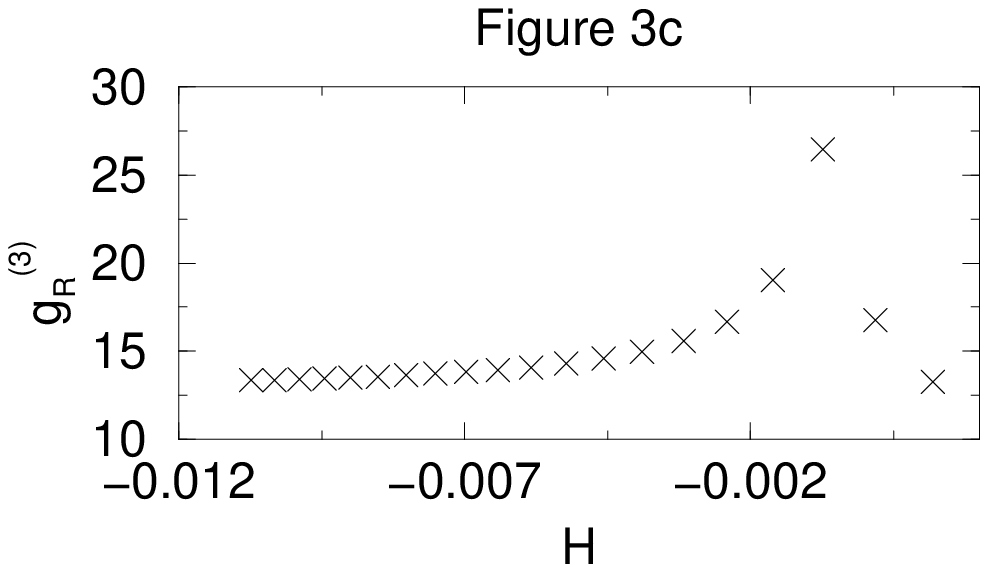,width=14cm} 
\end{figure}


\begin{thebibliography}{11}

\bibitem{ITZ-ZUB} 
C. Itzykson and J.B. Zuber,  Quantum Field Theory, McGrall- Hill ed., (1985), 
New York.
 
\bibitem{RAMOND} P. Ramond, {\it Field Theory: a Modern Primer},
Benjamin-Cummings, (1981).

\bibitem{STEFANOVICH}  E.V. Stefanovich, hep-th/0503076.

\bibitem{BAGAN} T. Barnes and G.I. Ghandhour, 
Phys. Rev. {\bf D22} (1980) 924.

\bibitem{STEVENSON} P.M. Stevenson, Phys. Rev. {\bf D32}, 1389 (1985).
   P.M. Stevenson, B. Alles and R. Tarrach, Phys. Rev. {\bf D 35},
2407 (1987).

\bibitem{BARMOSHE} W.A. Bardeen, M. Moshe, Phys. Rev. {\bf D 28}, 1372 (1983). 

\bibitem{KMV} A.K. Kerman, C. Martin and D. Vautherin, Phys. Rev. {\bf D47},
632 (1993).

\bibitem{THESE}
F.L. Braghin, Phys. Rev.   D 57, 3548 (1998).
 F.L. Braghin, Phys. Rev.  D 57, 6317 (1998).
F.L. Braghin, Doctoral Thesis,  1996.

\bibitem{COOPERetal} F. Cooper, S. Habib, Y. Kluger, E. Mottola, 
J.P. Paz, P.R. Anderson, Phys. Rev. {\bf D 50}, 2848 (1994).

\bibitem{DMITRASINOVIC} V. Dmitrasinovic, Phys. Lett. B 433, 362 (1998).

\bibitem{ZINNMOSHE}  M. Moshe and J. Zinn-Justin, hep-th/0306133.

\bibitem{FEYNMAN} R.P. Feynman in {\it Proceedings of the International
Workshop on Variational Calculations in Quantum Field Theory}, ed. by L. Polley
and D.E.L.Pottinger, Wangerooge, West Germany, september 1987, World
Scientific, Singapore (1988).


\bibitem{2gaussianas} G. Tiktopoulos, Phys. Rev. {\bf D 57}, 6429 (1998).

\bibitem{VASILEV-DAWSON} 
Oleg V. Vasil'ev and Kenneth A. Dawson, Phys. Rev. E 51, 765 (1995).

\bibitem{POSTGAU} F. Cooper, H. Shepard, C. Lucheroni, P. Sodano,
Physica {\bf D 68}, 344 (1993).
I. Stancu, Phys. Rev. {\bf D 43}, 1283 (1991).

\bibitem{KIM-YOU}
Chul Koo Kim and Sang Koo You, 
cond-mat/0212557 and references therein.

\bibitem{HADRON} J. Berges, in Proceedings of International Joint
Workshop on HADRON Physics and Relativistic Aspects
of Nuclear Physics, Angra dos Reis, RJ, Brazil, 
March-April (2004), Ed. by M. Bracco {\it et al},
AIP Proceedings 739, APS (2004).

\bibitem{OPTetc} 
M.B. Pinto, R.O. Ramos, Phys. Rev. {\bf D 60}, 105005 (1999).
D. Gromes, Zeit. fur Physik {C 71}, 347 (1996).
F. Cooper, L.M. Simmons, P. Sodano, Physica D 56, 68 (1992).

\bibitem{CEATEDESCO} P. Cea and L. Tedesco, Phys. Rev. D 55, 4967 (1997).

\bibitem{t-dep}  O.J.P. \'Eboli, R. Jackiw, S.-Y. Pi, Phys. Rev. {\bf D 37},
3557 (1988).


\bibitem{t-dep2}  F. Cooper, S. Habib, Y. Kluger, E. Mottola, Phys.
                     Rev. {\bf D 55}, 6471 (1997).       

\bibitem{t-dep3} 
Y. Tsue, A. Koike, N. Ikezi, hep-ph/0103246.
 S. Maedan, hep-ph/0412091.

\bibitem{LBetal} L.M.A. Bettencourt, K. Pao, J.G. Sanderson,
                 hep-ph/0104210.
Y. Bergner, L.M.A. Bettencourt, Phys. Rev. {\bf D 69} 045002 (2004).

\bibitem{BDVetal} D. Boyanovsky, H de Vega, R. Holman,
S.P. Kumar, R.D. Pisarski, Phys. Rev. {\bf D 5812} 5009 (1998).

\bibitem{BOYANOVSKY} D. Boyanovsky, M.D'Attanasio, H. de Vega, R. Holman,
                  D.-S. Lee, Phys. Rev. {\bf D 52}, 6805 (1995).


\bibitem{t-dep4} 
C. Destri, E. Manfredini, hep-ph/0001177; hep-ph/0001178.
S. Juchem, W. Cassing, C. Greiner, Phys. Rev. {\bf D 69}, 025006
(2004).


\bibitem{FLB2001PRD}
 F.L. Braghin, F.S. Navarra, Phys. Lett.  508 B, 243 (2001).
 F.L. Braghin, 
Phys. Rev. {\bf D 64} 125001 (2001).


\bibitem{ZINNJUSTIN}  J. Zinn Justin, {\it Quantum Field Theory and 
                    Critical Phenomena}, Oxford University
                   Press, Oxford (1996). 


\bibitem{HUMAPO} K. Huang, E. Manousakis and J. Polonyi, Phys. Rev. {\bf D35}
3187 (1987).

\bibitem{BRANCHINA}  V. Branchina, P. Castorina, M. Consoli, 
D. Zappal\`a, Phys. Rev. {\bf D 42} (1990) 3587.

\bibitem{TRACEANOMALY} For example: S.D. 
Joglekar and A. Mishra, Phys. Rev. {D 40}, 444 (1989).


\bibitem{CONSOLI} P. Cea, M. Consoli, L. Cosmai, hep-lat/0501013.

\bibitem{ASYMPFREE} C.M. Bender, K.A. Milton, Van M. Savage, 
Phys. Rev. {\bf D 62}, 085001 (2000).

\bibitem{FROLICH} J. Fr\"ohlich, Nucl. Phys. {\bf B200}, 281 (1982).

\bibitem{COSMO} 
D.H. Lyth and A. Riotto, Phys. Rep. {\bf 314}, 1 (1999). 
J.S. Heyl and A. Loeb, Phys. Rev. Lett. 88, 121302 (2002).




\bibitem{KERMANVAUTHERIN} A.K. Kerman, D. Vautherin, Ann. Phys. (N.Y.) 
{\bf 192}, 408 (1989). 


\bibitem{SCHRODINGER}  R. Jackiw, {\it in} {\em Field Theory and 
Particle Physics-
V J.A.Swieca Summer School}, eds. O. \'Eboli, M. Gomes, A. Santoro (World
Scientific, Singapore, 1990). 



\end{thebibliography}
\end{document}